# MR-based quantitative measurement of human soft tissue internal strains for pressure ulcer prevention

**Authors:**
**Alessio Trebbi***
Univ. Grenoble Alpes, CNRS, TIMC, 38000 Grenoble, France.
Alessio.Trebbi@univ-grenoble-alpes.fr
**Ekaterina Mukhina**
Univ. Grenoble Alpes, CNRS, TIMC, 38000 Grenoble, France.
ekaterina.mukhina@univ-grenoble-alpes.fr
**Pierre-Yves Rohan**
Institut de Biomécanique Humaine Georges Charpak, Arts et Métiers ParisTech, 151 bd de l'Hôpital, 75013. Paris, France
Pierre-Yves.ROHAN@ensam.eu
**Nathanaël Connesson**
Univ. Grenoble Alpes, CNRS, TIMC, 38000 Grenoble, France.
nathanael.connesson@univ-grenoble-alpes.fr
**Mathieu Bailet**
TwInsight, 38000 Grenoble, France
mathieu.bailet@twinsight-medical.com
**Antoine Perrier**
Univ. Grenoble Alpes, CNRS, TIMC, 38000 Grenoble, France
Groupe hospitalier Diaconesses–Croix Saint-Simon, 75020 Paris, France
TwInsight, 38000 Grenoble, France
perrier.antoine@gmail.com
**Yohan Payan**
Univ. Grenoble Alpes, CNRS, TIMC, 38000 Grenoble, France
Yohan.Payan@univ-grenoble-alpes.fr

*Corresponding author

## Abstract

Pressure ulcers are a severe disease affecting patients that are bedridden or in a wheelchair bound for long periods of time. These wounds can develop in the deep layers of the skin of specific parts of the body, mostly on heels or sacrum, making them hard to detect in their early stages.
Strain levels have been identified as a direct danger indicator for triggering pressure ulcers. Prevention could be possible with the implementation of subject-specific Finite Element (FE) models. However, generation and validation of such FE models is a complex task, and the current implemented techniques offer only a partial solution of the entire problem considering only external displacements and pressures, or cadaveric samples.
In this paper, we propose an *in vivo* solution based on the 3D non-rigid registration between two Magnetic Resonance (MR) images, one in an unloaded configuration and the other deformed by means of a plate or an indenter. From the results of the image registration, the displacement field and subsequent strain maps for the soft tissues were computed. An extensive study, considering different cases (on heel pad and sacrum regions) was performed to evaluate the reproducibility and accuracy of the results obtained with this methodology.
The implemented technique can give insight for several applications. It adds a useful tool for better understanding the propagation of deformations in the heel soft tissues that could generate pressure ulcers. This methodology can be used to obtain data on the material properties of the soft tissues to define constitutive laws for FE simulations and finally it offers a promising technique for validating FE models.

## 1. Introduction

Pressure ulcers are serious injuries generated by prolonged mechanical loadings applied on soft tissues. Most of the pressure ulcers occur on the heel and on the sacrum as these locations are loaded when patients are bedridden or wheelchair bound for long periods of time [1][2][3]. Ulceration requires high amounts of resources from the nursing cares and time to be healed and therefore represents a serious problem to the individual and the health care system[4]. In the worst cases, these complications lead to amputations and death. Depending on the type of external mechanical load, anatomy and tissue integrity, pressure ulcers can start superficially or deep within the soft tissues. Superficial wounds are formed on the skin surface and progress downwards, making them easy to identify in the early stages with solutions that can be promptly adopted to stop their progression. On the other hand, deep tissue injuries arise in muscle or fat layers around bony prominences and are often caused by high strains of the biological tissues. A value of 0.65 for the Green Lagrange (GL) maximal shear strain was provided by Ceelen et al. as a threshold that should not be exceeded to avoid any pressure ulcer [5]. This last case represents a major threat due to the impossibility to quickly identify the ulcer formation and promptly take action [6]. For this purpose, techniques to monitor the level of strain in the deep layers of the skin and underlying soft tissues are currently extensively investigated in the literature [7].

A common methodology to estimate internal tissue strains relies on FE modeling, with simulations that reproduce the body part morphology, tissue biomechanical parameters and the type of loading [8][9][10]. However, validation of FE simulations of the mechanical response of *in vivo* biological tissues to external mechanical loads has always been problematic. Keenan et al. report that none of the current heel models have been properly validated against independent experimental measurements and that further work is needed to develop models that are well validated to draw reliable clinical conclusions [8]. Regarding the buttock region, Savonnet et al. reached a similar conclusion stating that only few models were validated with experimental observations [9]. Because direct validation of internal mechanical strains is a challenging problem, many research works proposed to evaluate FE models of the foot in terms of their capacity to predict interface plantar pressure by comparing the contact pressure predicted by the FE model with the measurements from pressure mattresses [11]. Yet, as observed in Macron et al. [12] on data from 13 healthy volunteers, interface pressure distributions do not correlate with internal strains and one cannot be used to predict the other. This issue was partially addressed by Linder-Granz et al. [13] for a buttock FE model in a study where the authors compared contours of the computational domain in the deformed configuration predicted by the simulations to the ground truth segmented contours obtained from MR images. This comparison, however, considers only the external shape and not the quantity of interest, which is the local internal tissue displacement and associated tissue strains.

In an original contribution, Stekelenburg et al. [14] proposed to use MR tagging and phase contrast sequences on a rat leg model under indentation to assess local tissue displacements and compute the associated tissue strains. The main restriction of this approach is that the indenter (used inside the MR machine to deform the tissue) has to be applied rapidly and repetitively as the tagging grid fades within 1 s because of MR relaxation. This requirement can be complex to overcome with an MR compatible device. Moreover, this constraint does not allow for conventional control systems for the application of loads such as gravity, hydrostatic pressure or compression springs [15][16][17]. Additionally, with dynamic loads applied, the viscoelastic properties of the biological tissues could have an impact on the mechanical response, thus increasing the complexity to estimate the tissues passive mechanical properties from the experimental measurements.

Digital Volume Correlation (DVC) is an emerging non-invasive technique that allows to characterize experimentally material mechanical response to external loadings by tracking the displacement of natural patterns. From the displacement field, local strains can be computed. Combined with 3D MR images, DVC can, for example, be used to estimate human tissue internal strains [18]. From two MR datasets, one collected in an arbitrary undeformed configuration and another in a deformed configuration, the non-linear transformations that will align the MR volume at rest to the deformed one can be computed using a procedure call Image Registration. To illustrate the process, a graphical summary of the procedure is proposed based on data collected by the authors on the foot (Figure 1). DVC has previously been used in for *in vivo* strain estimation in human intervertebral discs, brain and leg muscles under external mechanical loading [19][20][21].

Our group has recently developed an MR-compatible device for applying controlled shearing and normal loads to the human heel pad [16]. With such a device, 3D MR volumes of the heel pad soft tissue can be imaged under various loads applied on the foot sole. This paper aims at describing the methodology proposed by our group to implement DVC on human soft tissues and at estimating the internal strains from the DVC-derived 3D displacement field. The long-term objective is to validate a FE model, in terms of its capacity to predict the localization and the intensity of the strain field in the soft tissues.

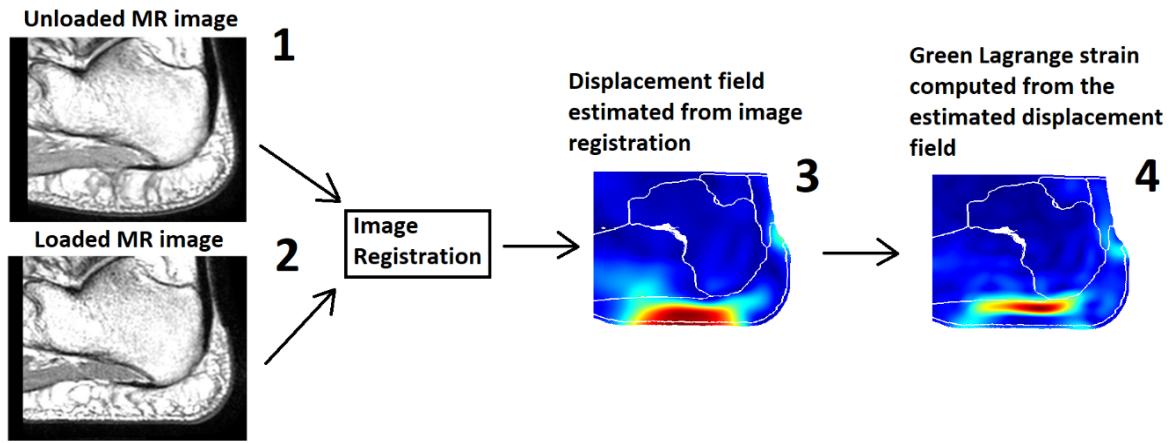

Figure 1 : Scheme of quantitative measurement of soft tissue internal strains obtained from image registration. Image 1: unloaded configuration. Image 2: Loaded configuration. The image registration estimates the displacement field (Image 3) that transforms the unloaded image into the loaded configuration. The strain field can then be derived from the displacement field (Image 4).

## 2. Materials and Methods

### 2.1. Materials: heel and sacrum MR datasets previously collected on one healthy volunteer

The MRI datasets used in this study have been collected in a previous study [17]. For the sake of clarity, the main details regarding the experimental setup, protocol and participant are summarized in the following paragraph. For more details, the reader is referred to the associated publication.

A healthy volunteer (male, 40 years old) gave his informed consent to participate in the experimental part of a pilot study approved by an ethical committee (MammoBio MAP-VS pilot study N°ID RCB 2012-A00340-43, IRMaGe platform, Univ. Grenoble Alpes).

For the heel MR image datasets, the volunteer was placed in a supine position with his right foot locked in a MR compatible device designed to apply both a normal force (15 N) or a combined normal-and-shearing force (15 N normal + 45 N shearing) on the heel pad by means of an indenting platform. The setup is illustrated in Figure 2A. A proton density MR sequence was used to collect 3D images composed of 512 x 428 x 512 voxels with voxel size of 0.3125 mm x 0.375 mm x 0.3125 mm (MRI system Achieva 3.0T dStream Philips Healthcare). Two acquisitions of the same unloaded configuration allowed to avoid having the same noise pattern between equivalent images in the subsequent image registration process in order to test the repeatability of strain calculation.

For the sacrum images, the subject was placed in the MR bed in a prone position. An indenter actuated by gravity applied a normal load (12 N) on the sacrum region. The 3D images were composed by 800×800×240 voxels with a dimension of 0.5 × 0.5 × 0.5 mm Figure 2B. Likewise, two acquisitions were collected in the unloaded configuration to test the repeatability of strain calculation.

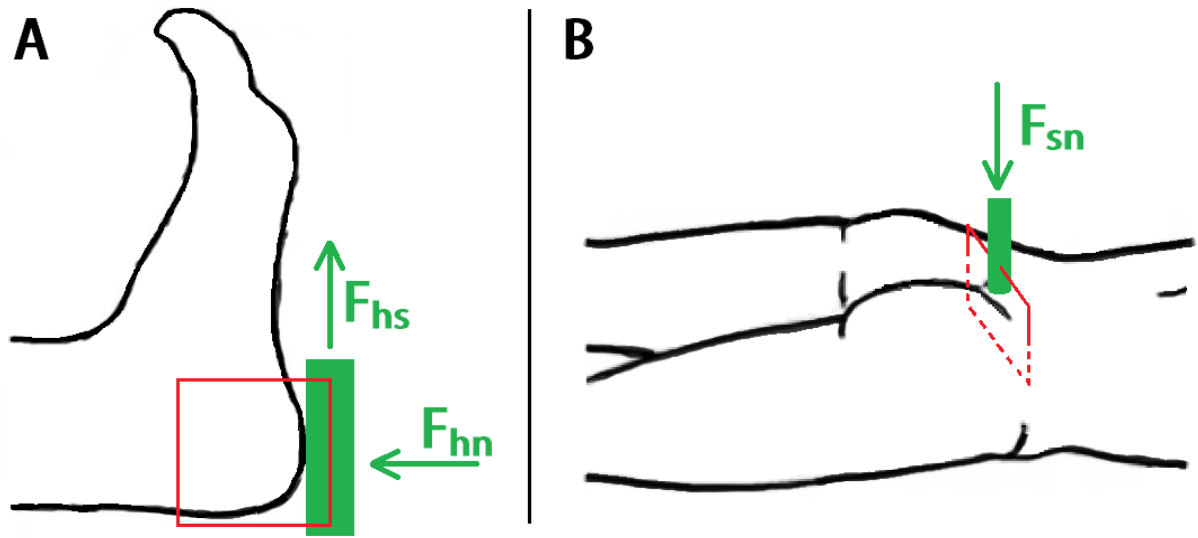

Figure 2 : (A) Scheme of the heel configurations during the MR acquisitions. The green rectangle represents the plate applying the loads. Direction of the loads is represented by the green arrows Fhn (Force heel normal) and Fhs (Force heel shear). The red rectangle shows the orientation of the MRI slice that will be shown in the rest of the paper. (B) Scheme for the sacrum configuration (Analogous to A). The green block represents the indenter with the respective Fsn (Force sacrum normal) applied. The indenter has the external shape of an ultrasound probe, 10-2 linear probe transducer developed by (Aixplorer, SuperSonic Imagine, France).

Four 3D MR images of the heel and three 3D MR images of the sacrum region were considered in this contribution and were referred to using a unique name as listed in Table 1.

| Name | Description | Load |
|---|---|---|
| Heel 01 | Unloaded heel – Acquisition 1 | 0 N |
| Heel 02 | Unloaded heel – Acquisition 2 | 0 N |
| Heel 1 | Heel with normal load | 15 N normal |
| Heel 2 | Heel with normal and shearing load | 15 N normal and 45 N shear |
| Sacrum 01 | Unloaded Sacrum – Acquisition 1 | 0 N |
| Sacrum 02 | Unloaded Sacrum – Acquisition 2 | 0 N |
| Sacrum 1 | Sacrum with normal load | 12 N normal |

Table 1 : List of MR acquisitions. The first name indicates the body location of the image. The unloaded configurations are indicated by the initial number 0 (01, 02). The loaded configurations are indicated by the integer positive numbers (1,2).

A 2D snapshot of each MR volume (presented as the red rectangle in Figure 2) is provided in Figure 3.

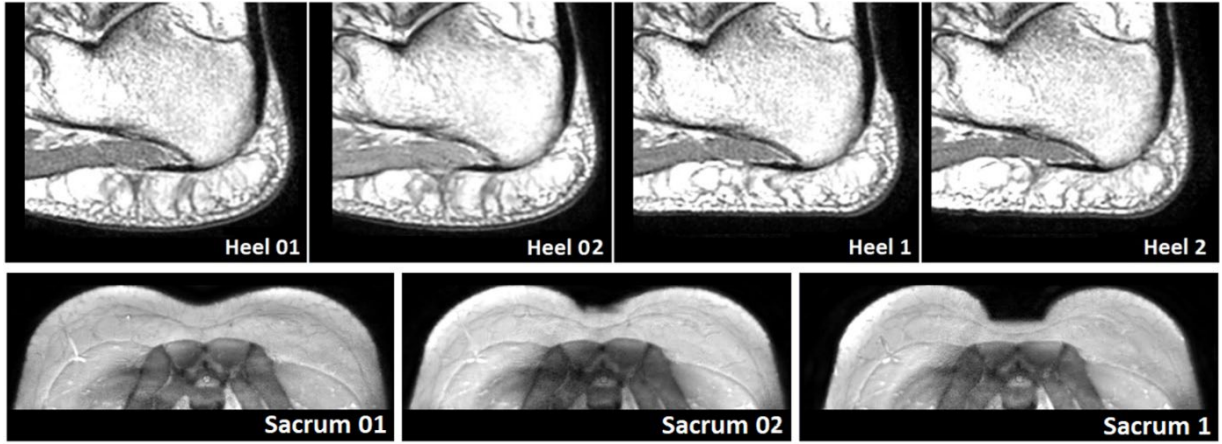

Figure 3 : Slices of the heel and the sacrum unloaded and loaded configurations described in Table 1. The respective region is indicated in Figure 2 by the red rectangle.

### 2.2. Rigid registration

The four MR volumes of the heel and the three MR volumes of the sacrum were rigidly registered to align the calcaneus bone and the sacrum bone respectively using the publicly available registration package Elastix [22].

### 2.3. Digital Volume Correlation between the loaded and the unloaded MR images

The registration package Elastix [22] was then used to perform DVC. Two images are involved in this registration process: the reference image $I_0(x)$ (unloaded configuration: Heel/Sacrum 01/02, called "fixed image" in the Elastix library) and the deformed image, $I_Q(x)$ (loaded configuration: Heel 1 and 2 and Sacrum 1, called "moving image" in the Elastix library), where $x$ represents the position of a point in the images. The registration between these two images defines a non-rigid deformation field $u_Q(x)$, which describes how the reference unloaded image transforms into the deformed image. Applying the deformation field to the reference image creates a transformed-deformed image $I_0\left(x + u_Q(x)\right)$ that aims to look identical to the deformed image.

The optimal deformation field was estimated by minimizing a cost function by means of an iterative optimization method (adaptive stochastic gradient descent) embedded in a hierarchical (multiresolution) scheme. The cost function relates to the similarity between the two images (*i.e.* the reference image and its transformation) using image features and was based on the Normalized Correlation Coefficient (NCC).

During the optimization step, the value of the cost function was evaluated at non-voxel positions, for which intensity interpolation with cubic B-Spline was used.

### 2.4. Computing mechanical strains from the DVC-derived displacement fields

From the displacement fields obtained by the registrations, strain maps were calculated as follows: The relation between the position $X$ of a material point in the undeformed configuration and its position $x$ in a deformed configuration $Q$ is described by the spatial displacement vector $u_Q(x)$ which consists of 3 components $u_{Qx}, u_{Qy}, u_{Qz}$ :

$$u_Q(x) = \left[u_{Qx}, u_{Qy}, u_{Qz}\right]^T \tag{1}$$

From these, the deformation gradient $F$ can be computed:

$$F = I + \frac{\partial u}{\partial X} \quad (2)$$

And the right Cauchy-Green deformation tensor $C$ deduced:

$$C = F^T F \quad (3)$$

The Green Lagrange principal strains:

$$E_p = eig(\frac{1}{2}(C - I)) \quad (4)$$

The maximum GL shear strains are defined as:

$$E_s = \frac{1}{2} * \max(|E_1 - E_2|, |E_1 - E_3|, |E_2 - E_3|) \quad (5)$$

### 2.5. Uncertainty of the Image registration procedure

To evaluate the uncertainty of the DVC we consider six evaluation Cases A to F. The first three cases are related to the repetition of the same strain measurement and to the analysis of the differences between the respective results (reproducibility of the registration). The last three cases focus on the ability of DVC to estimate a known a priory strain field (accuracy of the registration).

#### 2.5.1. Reproducibility

Reproducibility refers to the closeness of agreement between test results. In this section, we propose to evaluate the reproducibility of strain calculation through image registration. Two acquisitions of the unloaded configurations of the heel and sacrum (namely Heel 01 and Heel 02 and sacrum 01 and sacrum 02 respectively) were registered to the same moving image (Heel 1 and Sacrum 1 respectively). The corresponding strain maps are computed from the two estimated deformation fields. The reproducibility is then inspected by analyzing the differences between these two strain maps. Three cases, summarized in Table 2, are considered: heel under normal load (A), heel under normal+shearing load (B) and sacrum under normal load (C).

| Fixed image | Moving image | Case | Displacement field | Shear strain field |
|---|---|---|---|---|
| Heel 01 | Heel 1 | A | $DA_{011}$ | $SA_{011}$ |
| Heel 02 | Heel 1 | | $DA_{021}$ | $SA_{021}$ |
| Heel 01 | Heel 2 | B | $DB_{012}$ | $SB_{012}$ |
| Heel 02 | Heel 2 | | $DB_{022}$ | $SB_{022}$ |
| Sacrum 01 | Sacrum 1 | C | $DC_{011}$ | $SC_{011}$ |
| Sacrum 02 | Sacrum 1 | | $DC_{021}$ | $SC_{021}$ |

Table 2 : List of image registrations to evaluate the reproducibility of strain calculation from image registration. Each line represents an image registration composed by its fixed and moving image. The tests are grouped in three Cases: A) Heel with normal load, B) Heel with normal+shearing load, C) Sacrum with normal load. The resulting displacement fields and shearing strain field are respectively denoted with the letters D and S. The second letter in the field nomenclature reports the respective case of the registration. The numbers report the name of the fixed and moving images.

#### 2.5.2. Accuracy

Accuracy reflects how close a data is to a known or accepted value. In this section, we propose to evaluate the accuracy of our image registration procedure to identify a known *a priori* strain field. We focus specifically here on the images of the heel. Two different displacement fields are considered:

1. For the first case, an artificial displacement field $D_{FEM}$ is generated from a Finite Element (FE) simulation. A rectangular parallelepiped volume with the same size of the 3D MR images is first generated in ANSYS 19.2 APDL (ANSYS, Inc., Canonsburg, PA). This volume is then meshed with 8-nodes hexahedral elements and a linear elastic material model is implemented. The 3D mesh is composed of 24389 hexahedral elements. The nodes on the sides of the parallelepiped are fixed in order to avoid any displacements outside of the defined volume. A set of 196 internal central nodes located on the same XZ plane are then submitted to a prescribed displacement boundary condition in a normal (Y) and in a tangent (X) direction (Figure 4). The displacement field computed by ANSYS is then extracted for all the nodes of the parallelepiped and interpolated to fit the resolution of the MR images. The corresponding displacement field is applied to the unloaded image (Heel 01) to generate a new artificially loaded image of the heel, named Heel FEM (Table 3 and Figure 5). It is worth noting that the objective of the FE method is mainly to produce a known *a priori* displacement filed. This displacement filed will be subsequently estimated through the image registration technique. The simulation itself, and the artificially generated image Heel FEM, do not have any real physical meaning. The main benefit of using such an FE solver is the possibility to get a ground-truth strain field that can be compared to values estimated from image registration.

2. For the second case, the previously computed displacement field $DA_{011}$ is applied to the unloaded image (Heel 01) to generate a new artificially loaded image of the heel, named Heel TRA (from the word transformed) (Table 3 and Figure 5).

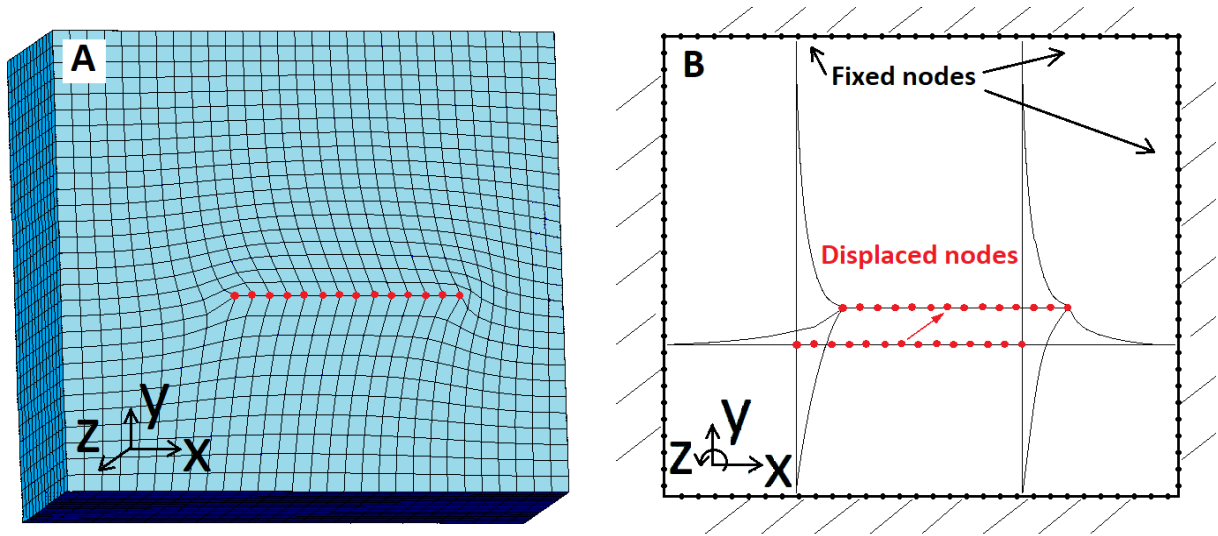

Figure 4 : Generation of an artificial displacement field from a FE simulation generated by Ansys. The size of the cube matches with the size of the MR images of the heel. A selection of nodes (red dots) on a plane orthogonal to the y axis was displaced as boundary conditions. (A) Section of the simulated cube along a plane orthogonal to the z axis. (B) Schematization of the boundary conditions imposed. The external nodes were fixed, and the selection of red nodes was displaced.

| Image   | Applied displacement field | Artificial image | Shear strain field |
|---------|----------------------------|------------------|--------------------|
| Heel 01 | $D_{FEM}$                  | Heel FEM         | $S_{FEM}$          |
| Heel 01 | $DA_{011}$                 | Heel TRA         | $SA_{011}$         |

Table 3 : List of transformations to create the artificial images to test the accuracy of strain calculation through image registration. The image column lists the images to be transformed. The displacement field column lists the transformation to be applied to generate the artificially deformed image.

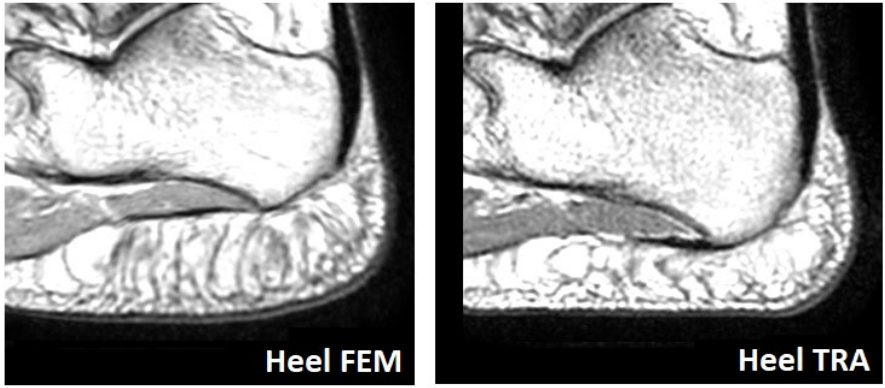

Figure 5 : Artificial images obtained once the displacement fields $D_{FEM}$ and $DA_{011}$ are applied to the unloaded image Heel 01.

Image registration was then computed between Heel 02 and the two artificially deformed images Heel FEM and Heel TRA. Note here that having two acquisitions of the same unloaded configuration (Heel 01 and Heel 02) allowed to implement different noise patterns during the registration process in the fixed and moving image (Cases D and F of Table 4). On the other hand, to show the impact of having the same noise pattern between the fixed and the moving image the image Heel 01 was also considered for Case E (table 4).

| Fixed | Moving | Case | Displacement field | Shear strain field |
|---|---|---|---|---|
| Heel 02 | Heel FEM | D | DD | SD |
| Heel 01 | Heel FEM | E | DE | SE |
| Heel 02 | Heel TRA | F | DF | SF |

Table 4 : Following cases A, B and C mentioned in table 2, cases D, E and F relate to the estimation of the accuracy of strain calculation through image registration. The shear strain fields SD and SE will be compared with $S_{FEM}$. The shear strain fields SF will be compared with $SA_{011}$.

### 2.5.3. Error quantification

The error estimation was performed analyzing the obtained strain fields with a Bland–Altman plot. This representation is a method of data plotting used in analyzing the agreement between two different set of data corresponding to the same measurement. The plotted graph shows the error distribution throughout the whole range of measured strain values.

## 3. Results
### 3.1. Strain measurements for heel under normal load (case A, Table 2)

The distribution of the DVC-derived displacement field in the heel domain under normal load is given in the sagittal slice containing the highest shear strains (Figure 6A). The highest displacements are uniform in the area where the plate was in contact with the plantar skin of the heel. Figure 6B shows the corresponding maximal GL shear strains computed from the displacement field. Shear strains are concentrated around the lower part of the calcaneus bone propagating towards the plantar fascia and the flexor digitorium brevis.

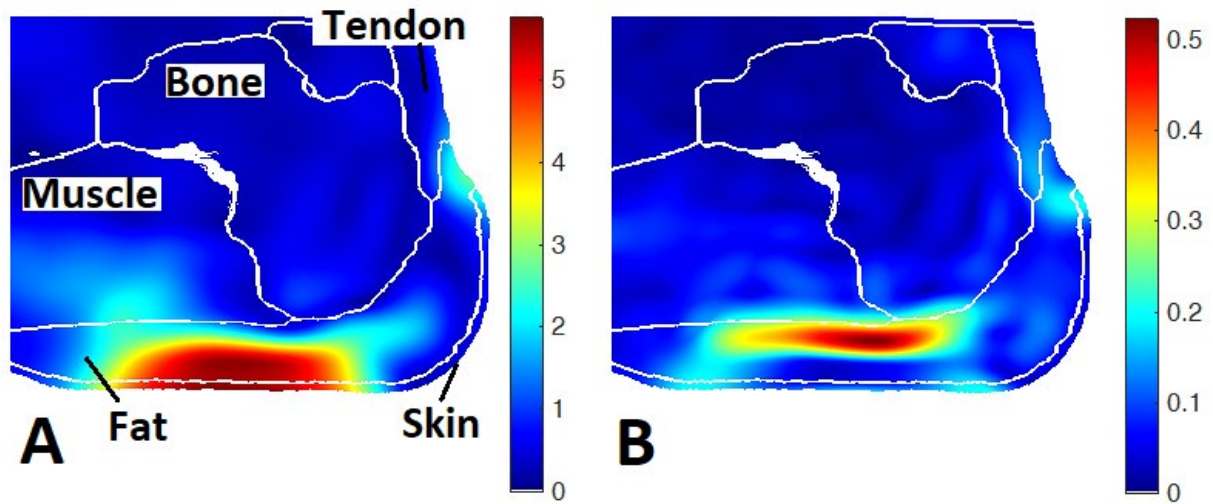

Figure 6 : Case A. Biological tissues are delimited by white lines. A slice from the MR volume is shown from the sagittal plane corresponding to the location of the highest shear strain. (A) Visual representation of $DA_{011}$. Modulus of displacement field [mm] for heel under normal load. (B) Visual representation of $SA_{011}$. Max GL shear strain field for heel under normal load (0.5 corresponds to 50% of deformation).

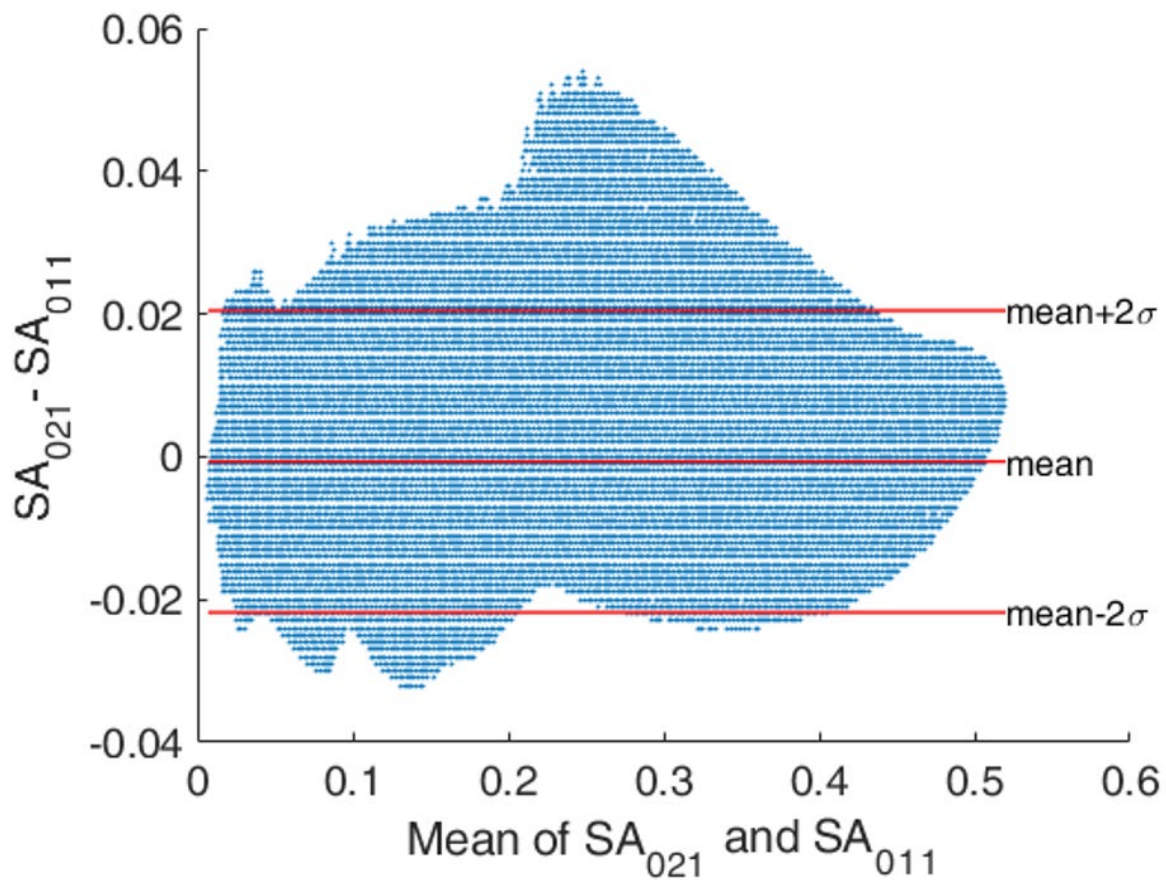

Figure 7 : Bland-Altman plot referring to the strain estimation computed from Case A: heel under normal load. The upper and lower red line correspond the 95% confidence interval, meaning that 95% of the values have an error lower than 0.02 strain. The most relevant part of the plot is the region with the highest values of the strains 0.4-0.5 as these can represent the threat for tissue damage.

The agreement between $SA_{011}$ and $SA_{021}$ was described graphically with a Bland-Altman plot (Figure 8) with mean of differences, reported with corresponding 95% confidence interval (CI), and lower and upper limits of agreement, calculated as mean ± 2$\sigma$ (where $\sigma$ represents the standard deviation SD). Differences were assessed using a Wilcoxon-Signed-Rank Test (paired data) at the default 5 % significance level.

**3.2. Strain measurements for heel under normal+shearing load (case B, Table 2)**

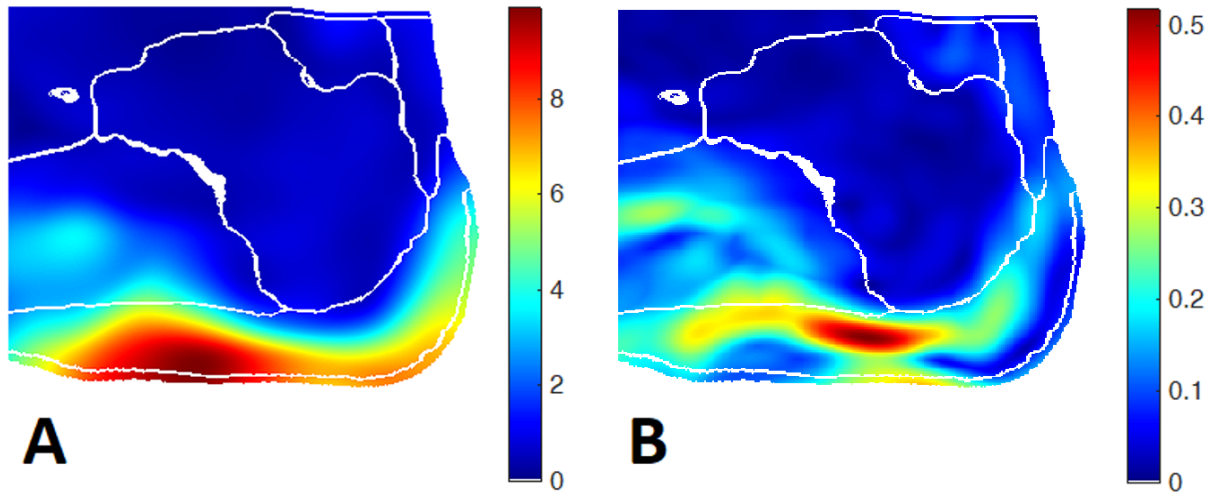

Figure 9 : (A) Modulus (in mm) of displacement field for heel under shearing load $DB_{012}$. (B) Max GL shear strain field for heel under shearing load $SB_{012}$.

The application of the shearing load had a relevant impact on the soft tissue displacements. The plate moved the posterior and the plantar regions of the heel skin towards the forefoot. This caused the shear strains to propagate on a wider region of the fat pad and the muscle (Figure 8). A concentration of high levels of strains is found in the fat pad under the flexor digitorum brevis.

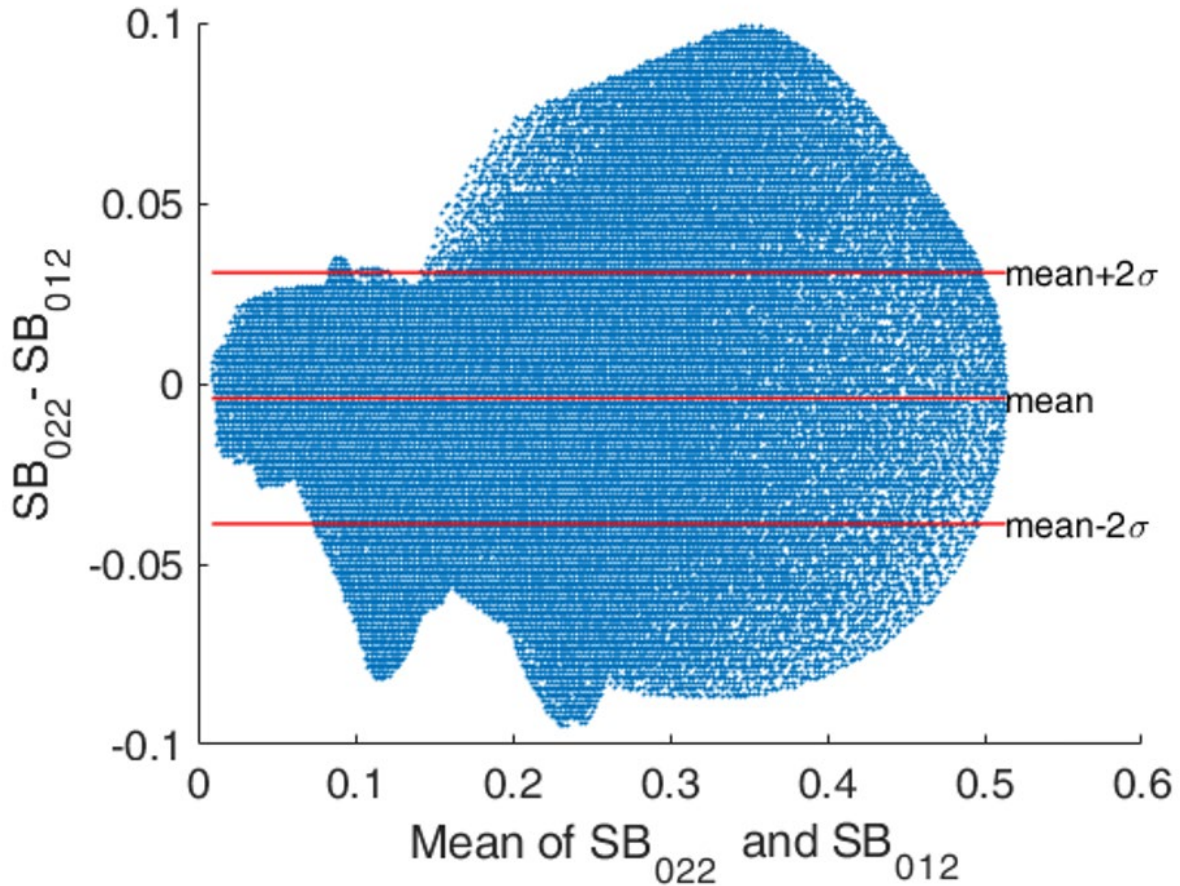

Figure 10 : Bland-Altman plot referring to the strain estimation computed from Case B: heel under normal+shearing load. Error magnitude is around twice higher than the configuration with normal load only (Figure 7).

Figure 9 shows the correlation between the strain measurements of the heel under normal+shearing loads (Case B of Table 2). Errors of 0.1 are observed across most of the strain intensities even for the highest strains (around 0.5). These errors tend to narrow down for the peak values. The SD shows that 95% of voxels have a strain error lower than 0.04. In general, this shearing configuration (Case B) shows errors with a double intensity and twice the propagation with respect to the normal load configuration (Case A).

### 3.3. Strain measurements for sacrum under normal load (case C, Table 2)

For the sacrum loading configuration (Figure 10), the highest levels of displacements are found around the edges of the indenter. Shear strains are concentrated on the soft tissues around the contact area between the indenter and the skin. Adipose tissue and skin are subject to the highest levels of strains.

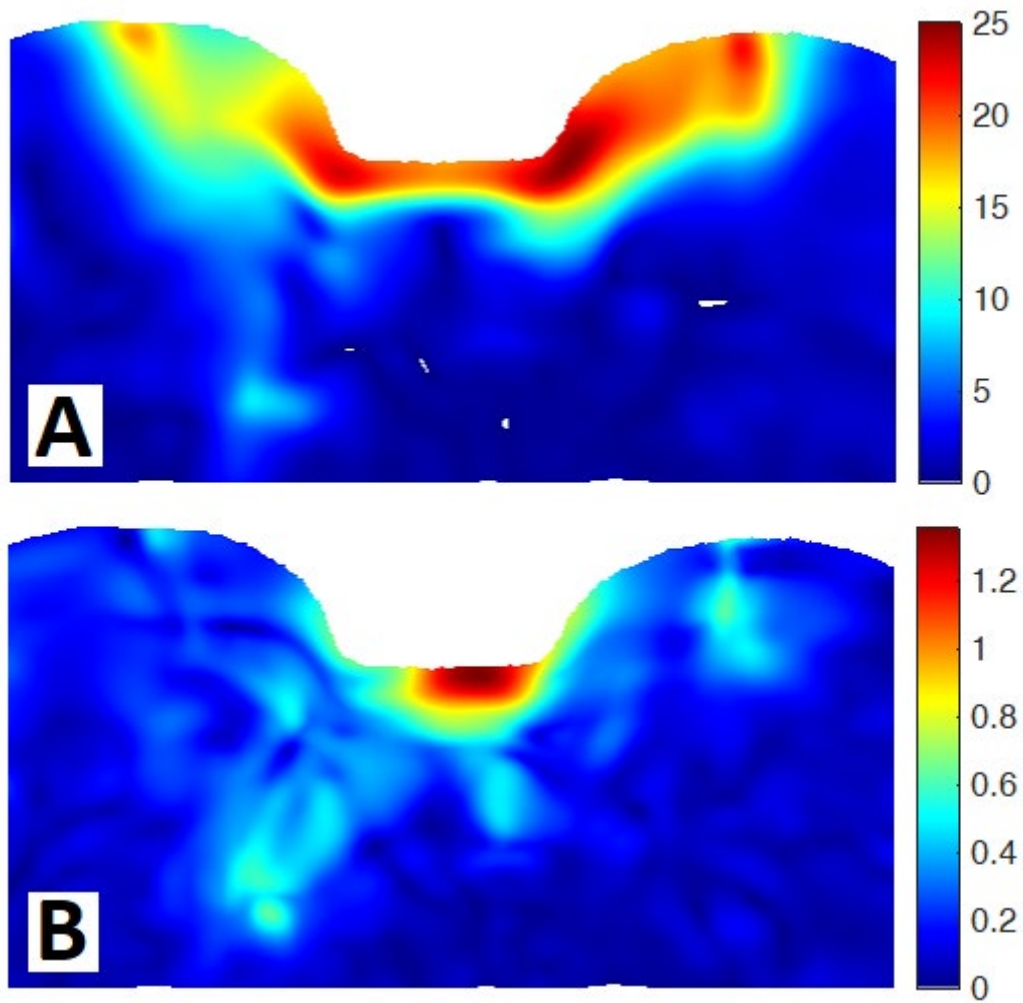

Figure 11 : (A) Modulus (in mm) of displacement field for sacrum under normal load $DC_{011}$. (B) Max GL shear strain field for sacrum under normal load $SC_{011}$.

Figure 11 presents the Bland-Altman plot between the shear strain measurements produced by an indenter on the sacrum region (Case C of Table 2). In this case, the errors are considerably higher than what was observed for the heel application. Errors of 0.3 are spread throughout the image and the SD describes an error distribution where 95% of the voxels have an error that is lower than 0.15.

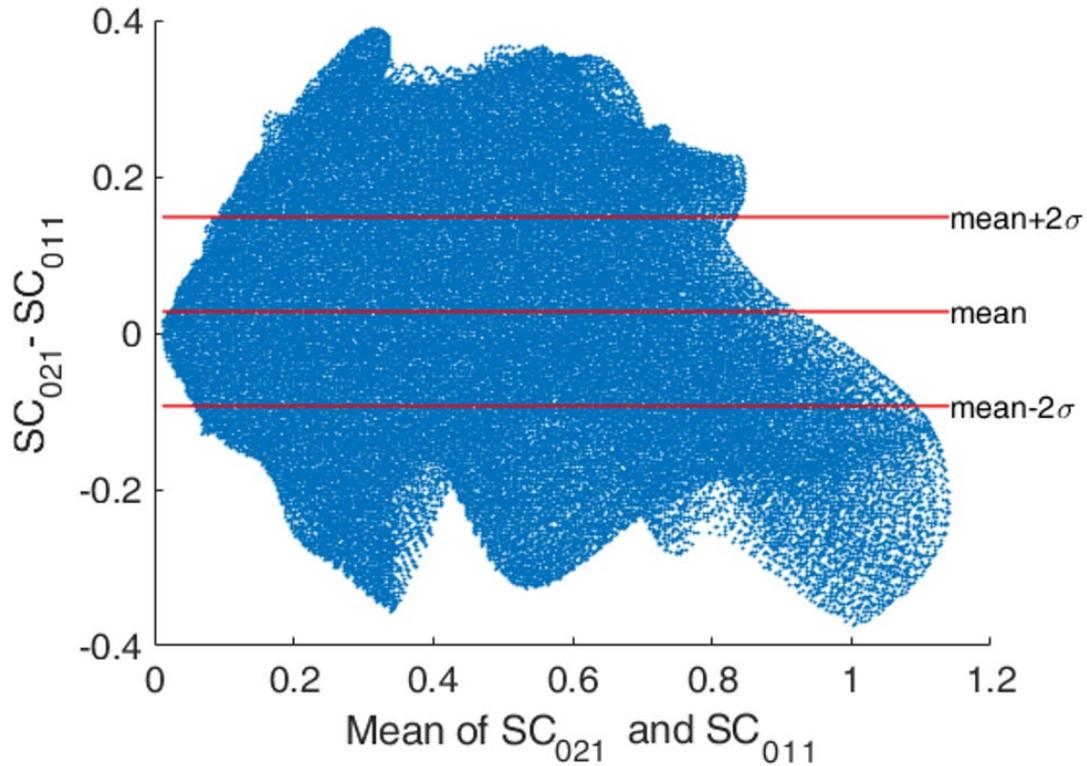

Figure 12 : Bland-Altman plot referring to the strain estimation computed from Case C: sacrum under normal load. Errors are considerably higher than the previous configurations of the heel.

### 3.4. Estimation of strain field generated by the FE model (case D, Table 4)

Figure 12 shows the results of image registration in the estimation of the artificial displacement field generated by Ansys (Figure 4). Magnitudes of displacements were selected in order to generate strains comparable with Cases A and B.

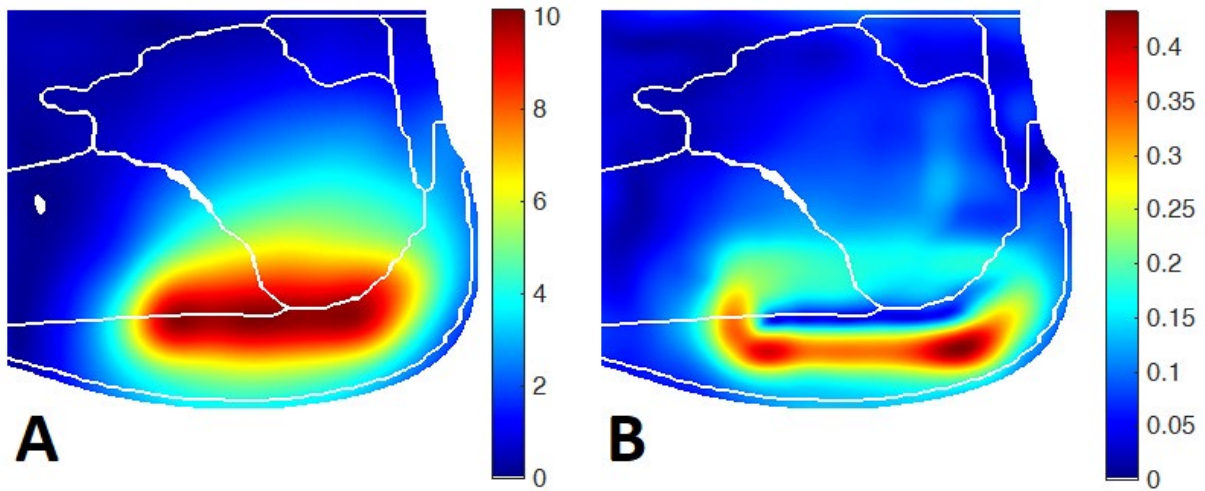

Figure 13 : (A) Estimation of the displacement field (mm) generated by Ansys DD. (B) Estimation of GL max shear strain generated by Ansys SD.

Figure 12 presents the correlation between the strain field calculated by Ansys and the corresponding measurements obtained by image registration (Case D of Table 4). The error distribution is comparable to Case A. For the regions with the highest levels of strains, the measurements slightly underestimate the strains since the points distribution shows an inclination that is higher than the red line. This artifact could be a result of the transformation step described in Table 3. In this process, some details of the original displacement field could have been lost in the image reconstruction after the application of the displacement to the respective voxels.

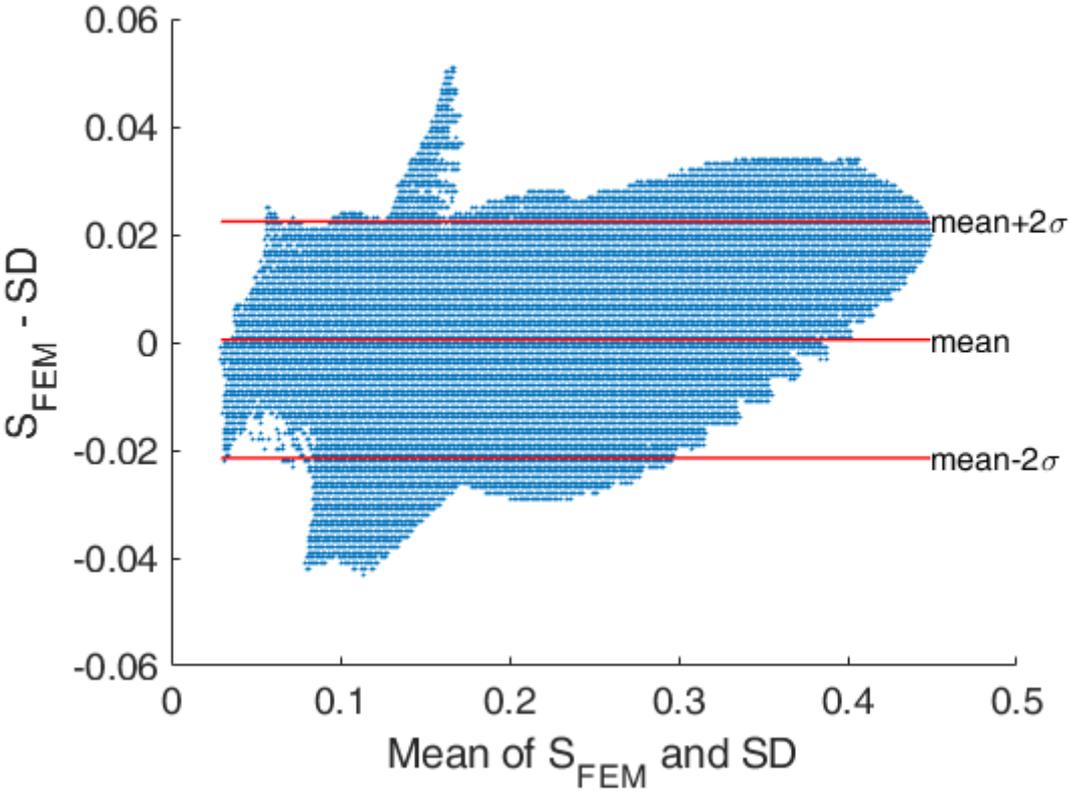

Figure 14 : Bland-Altman plot referring to the strain estimation computed from Case D: Displacement field generated by Ansys. The intensity of errors is around 0.02, which is comparable with case A (Figure 7).

**3.5. Deformation field from Ansys – Same noise pattern (case E, Table 4)**

This case is running the registration between two images with the same noise pattern, undeformed (Heel 01), and artificially deformed (Heel FEM). Using the same image helps considerably the algorithms of the image registration process since the noise pattern present in the unloaded image matches the one of the unloaded image. This allows to easily identify the respective deformation matching the voxels with their equivalent copy in the respective deformed image. Results in terms of error distribution are as expected very precise showing a relevant strain field estimation (Figure 14). This reflects the described facilitations in terms of using an image and its deformed version in the registration process.

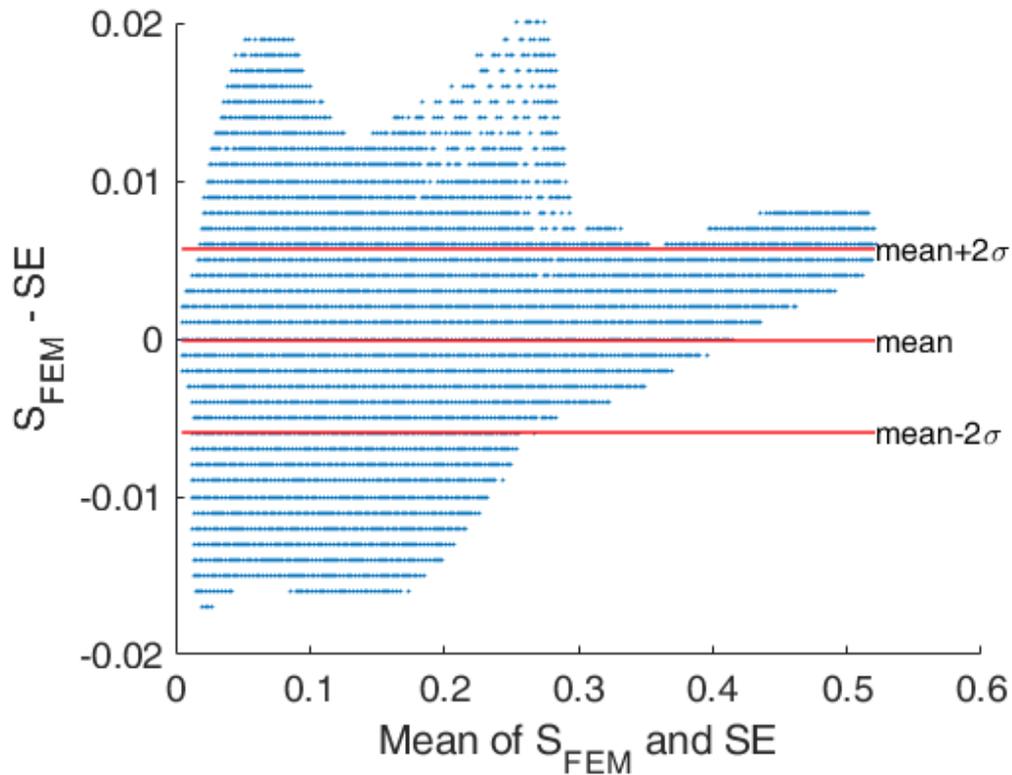

Figure 15 : Bland-Altman plot referring to the strain estimation computed from Case E: Displacement field generated by Ansys. Errors are lower than the other considered cases. This is due to the same noise pattern between the fixed and moving image in the registration procedure.

### 3.6. Deformation field from Elastix (case F, Table 4)

Case F is analogous to Case D with the main difference that the considered displacement field is not generated by Ansys but is taken from the image registration computed in Case A. The error distribution in terms of maximal error and SD is comparable to Case A (Figure 15). For the regions with the highest levels of strains, as detected also in case D, the measurements slightly underestimate the strains. In this case, the deformed image is also the result of an image transformation reported in Table 3.

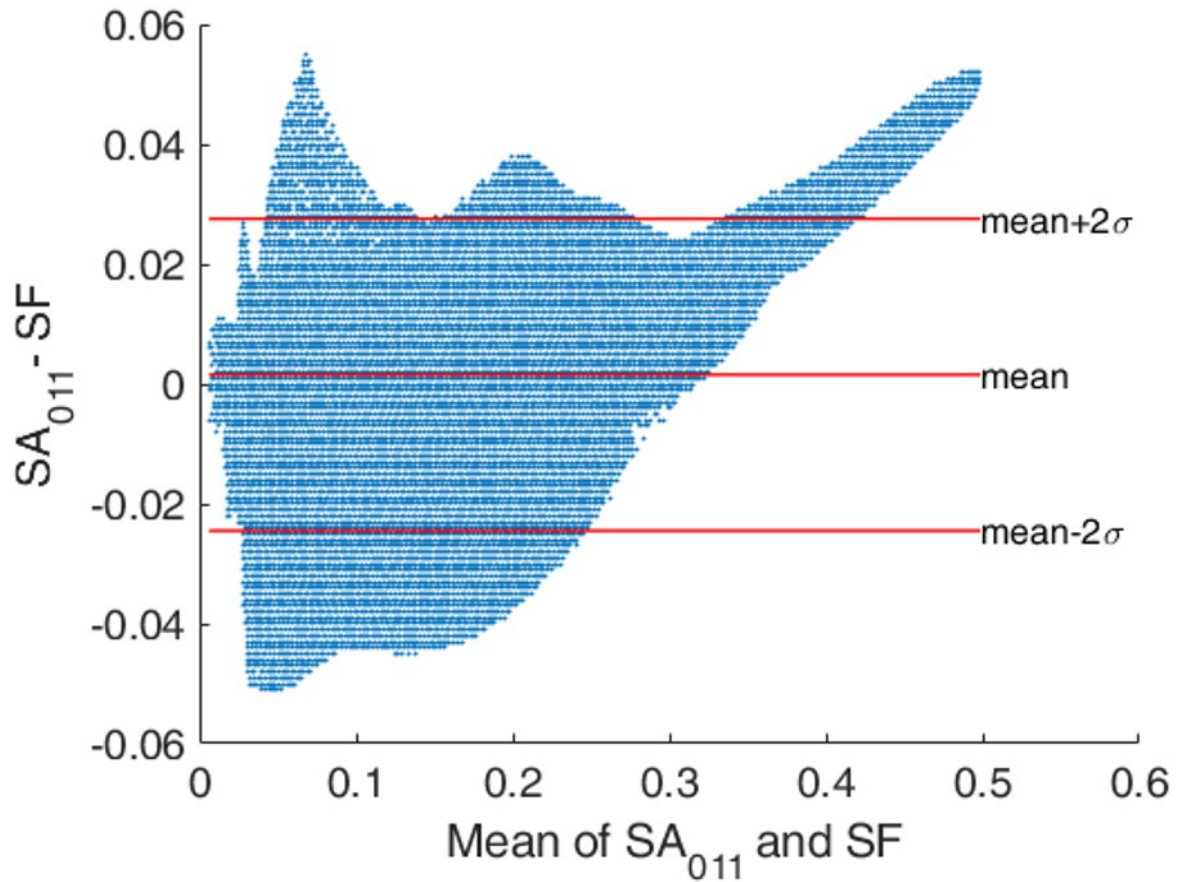

Figure 16 : Bland-Altman plot referring to the strain estimation computed from Case F: Displacement field generated by Elastix in Case A. The intensity of errors is around 0.02 comparable with Case A and D (resp. Figure 7 and Figure 13). This shows that, in the analyzed cases, similar images (Heel 01, 02, 1) generate errors of comparable magnitude (0.02).

## 4. Discussion

In this study, a method to estimate 3D internal tissue strains in the heel and sacrum regions based on DVC-derived displacement fields was developed in the context of pressure ulcers etiology. The methodology to implement DVC between two MR exams of human soft tissues (one at rest and the other one deformed) and to estimate the internal strains from the DVC 3D displacement field was first described. The implemented methodology requires a MR compatible device to apply loading on the skin surface during the acquisition of MR images. The obtained acquisitions were then used as input for 3D image registration. Images were first aligned based on the fixed body part (like bones) and then the non-rigid transformation was calculated. This transformation consists of a displacement field mapping every voxel between its initial position in the unloaded image and its final position in the loaded image. The GL shear strains are then computed from this displacement field. This methodology was implemented to analyze strain propagation in body regions that are critical in terms of pressure ulcer development: heel and sacrum.

For the results related to the heel pad, the calculated strain maps show as expected shear and compressive strain concentrations around the bony prominences of the calcaneus. This is coherent with the study of Luboz et al. which highlighted that strains generated around the calcaneus head strongly depended on the shape of this calcaneus bone [23]. Strains values in the deep tissues of the human heel were considerably higher than those in superficial tissue layers (Figure 6B). This is consistent with previous findings listing the strain concentration in deep tissues as a key aspect in the etiology of ulceration [24]. The strains were concentrated in the fat pad region and propagating towards the interface with the muscular region. Oomens et al. identified skeletal muscle and fat as the two main biological tissues where pressure ulcers could develop [25]. The application of a shearing load pushing the first layer of skin towards the forefoot generated significantly higher shearing loads in the anterior region of the fat pad compared with the configuration with only normal plantar pressure. Shear loadings can therefore impact significantly wider regions with higher levels of shear strains compared with normal loads of comparable intensity. This confirms what Ceelen et al. stated, namely (1) that shearing loads are more dangerous to treat than normal loadings, in terms of shear strain concentrations, and (2) that they must be taken into consideration for an effective pressure ulcer prevention [26].

For the results related to the sacrum, the calculated displacement and strain maps have values that are significantly higher than the results from the heel. The application of a load by means of an indentation device with a small contact area is probably more likely to generate higher shear strains right on the contact surface between the skin and the indenter [26].

The second objective of this article was to evaluate in a general way the reproducibility and the accuracy of strain calculation through image registration. Respectively, two main methodologies were presented: one related to the repetition of the same strain measurement from an equivalent set of images, and the other one to the calculation of a known *a priory* displacement field.

Concerning reproducibility as how much two equivalent measurement match, Figure 7, 9, 11 are considered. Comparing the strain error distribution between image registrations of the heel related to Case A and Case B, we found that errors are twice higher and more distributed in the case where the shearing load is applied. This suggests that strain measurement from image registration is affected by the type of deformation applied on the soft tissues. A possible explanation for this effect can be related to the fact that a normal load displaces the skin in a normal direction generating a clear displacement of the edge between the portion of image representing the biological tissues and the dark background (see Figure 3 Heel 01 and Heel 1). On the other hand, a shear load displaces the skin only in a tangent direction to the surface of the skin without generating any clear movement of the edge between the skin and the background (see Figure 3 Heel 1 and Heel 2).

The image registration related to the sacrum has a much wider strain error distribution and values compared to the examples of the heel. This implies that strain measurement from image registration strongly depends on the image characteristics. To explain the reasons behind this we can try to analyze how image registration works. The first steps in the algorithms of image registration are feature detection and feature matching [27]. Salient and distinctive objects as edges are considered as features. The accuracy of image registration therefore directly depends on the quality of the acquired images to define clearly these edges [28]. The main parameters that characterize the quality of digital images are related to resolution and noise [29]. Noise is generated by the statistical fluctuation of the value from voxel to voxel. A common measurement of noise is the standard deviation, a measure of how spread out the values of the pixels are. The lower the standard deviation, the higher the accuracy of the average voxel value [30]. Spatial resolution is the ability of the imaging system to detect small objects that are close to each other [31]. The size of the voxels defines the maximum spatial resolution. However, image resolution is also influenced by other parameters such as blur factors. The most common blur factor is motion blur: when motion occurs during acquisition, the boundaries of patient structures will move from their initial position, making the boundaries blurred in the image. The motion can in general be reduced by fixing the body part with heavy MR-compatible pillows or casts [16]. These solutions, however, are ineffective when motions are generated by physiological movements such as breathing, peristalsis or heart beats. The line spread function (LSF) can be used to evaluate and quantify spatial resolution [32][33]. From this parameter, it was calculated that in the more crucial region of the images, the MR image of the heel had a quality parameter related to the spatial resolution that was 4.5 time higher than the one calculated for the sacrum images. It is possible therefore that this aspect played a crucial role in the strain estimation through image registration, thus decreasing significantly its reproducibility.

Concerning accuracy as how close a measurement is to a known or accepted value, Cases D and F were considered. These cases report errors of shearing strains around 0.02. This value can be compared to the level of strains that is considered to be sufficient to generate significant tissue damage. According to Ceelen et al., this value is around 0.65 for the shearing strain [5].

An interesting aspect is related to Case E, which uses the same image Heel 01 and its transformed version (Heel TRA) for the estimation of the displacement field. In this case, the strain errors are much less distributed (with a 95% confidence value below 0.007 error). This is due to the fact that any variability due to noise, or other artifacts, present in both images will have an impact on the strain estimation. This implies that for images with appropriate quality levels, this methodology can reach high accuracy.

The diversity of results obtained between the heel and sacrum applications implies that crucial further research is therefore required in finding the relation between specific image quality parameters and the respective error distribution in the strain calculation. This would permit in fact to select the image acquisition protocols in order to obtain the type of images to minimize errors in the registration process.

An advantage of the proposed methodology to calculate strains is that no additional tool to perform the error estimation is required. Considering Case E, the error estimation can be performed just with an additional image transformation (Table 3) and the respective image registration.

It is clear that the accuracy of the results is strongly related to the image registration process and to the selected parameters to perform it. By tuning the respective parameters of the registration process, it is possible to identify smaller deformations or to select the amount of volume compression and expansion. An optimization for the selection of the ideal parameters of the registration for the related application will be considered in the future steps to improve the accuracy of this methodology.

It must be considered that this work was based on specific mechanical configurations of a single subject meaning that results obtained are to be considered specific to this application. Fat and muscle biomechanical properties can change significantly as a consequence of diseases (for example, diabetes) and chronic immobilizations [34][35]. This inter-subject variability may introduce significant variations in the strain calculations making imperative to analyze each subject specifically.

## 5.     Conclusion

The results obtained from the practical application on the heel and sacrum, in terms of location and magnitude of strains, are in line with the literature. This technique of calculating strains offers broad new possibilities to analyze the impact of external loads on the internal state of the soft tissues. The standard technique of FE is a very complex and time-consuming task involving segmentations, meshing and selections of proper constitutive laws. The possibility of strain calculation through image registration can provide results in terms of strain propagation in a significantly faster framework and offer the possibility for comparison and validation with results obtained from FE simulations. The present study proposed to quantify subdermal tissue strain distributions on the heel and sacrum from image registrations based on MR-acquisitions. This data is crucial for understanding the etiology of pressure ulcers that occur in the deep tissues of the heel pad.

The pilot study described here indicates that the crucial steps for computing strains from image registration are feasible to be implemented in a wider study. Further research will include analysis on more subjects and with different loading configurations, together with the adaptation of this methodology to different parts of the body to gain insight into the relative mechanical soft tissue properties.

## Acknowledgments

This research has received funding from the European Union's Horizon 2020 research and innovation programme under the Marie Skłodowska-Curie Grant Agreement No. 811965; project STINTS (Skin Tissue Integrity under Shear). IRMaGe MRI facility was partly funded by the French program "Investissement d'Avenir" run by the "Agence Nationale pour la Recherche"; grant "Infrastructure d'avenir en Biologie Sante" - ANR-11-INSB-0006.

## Conflict of interest

None

## Ethical approval

A volunteer (male, 40 years old) agreed to participate in an experiment part of a pilot study approved by an ethical committee (MammoBio MAP-VS pilot study). He gave his informed consent to the experimental procedure as required by the Helsinki declaration (1964) and the local Ethics Committee.